\documentclass[amsmath,aps, prl, reprint, superscriptaddress]{revtex4-2}
\usepackage{graphicx} 
\usepackage{braket}
\usepackage{physics}
\usepackage{amssymb}
\usepackage{mathtools}
\usepackage{hyperref}
\usepackage[capitalize]{cleveref}
\usepackage{microtype}
\usepackage{float}
\usepackage{xcolor}
\usepackage[normalem]{ulem}

\begin{document}

\title{Postselected Criticality in Measurement-Induced Phase Transitions}
\author{Dolly Nambi}
\thanks{These authors contributed equally to this work.}
\affiliation{Department of Physics and Astronomy, Louisiana State University, Baton Rouge, Louisiana 70803, USA}

\author{Kabir Khanna}
\thanks{These authors contributed equally to this work.}
\affiliation{Department of Theoretical Physics, University of Geneva, 1211 Geneva, Switzerland}
\affiliation{Department of Physics, University of Massachusetts Amherst, Amherst, Massachusetts 01003, USA}
\affiliation{Geneva Quantum Center, University of Geneva, Geneva, Switzerland}

\author{Andrew Allocca}
\affiliation{Department of Physics, City College of New York, City University of New York, New York, New York 10031, USA}
\affiliation{Department of Physics and Astronomy, Louisiana State University, Baton Rouge, Louisiana 70803, USA}

\author{Thomas Iadecola}
\affiliation{Department of Physics, The Pennsylvania State University, University Park, Pennsylvania 16802, USA}
\affiliation{Center for Theory of Emergent Quantum Matter, The Pennsylvania State University, University Park, Pennsylvania 16802, USA}
\affiliation{Institute for Computational and Data Sciences, The Pennsylvania State University, University Park, Pennsylvania 16802, USA}
\affiliation{Materials Research Institute, The Pennsylvania State University, University Park, Pennsylvania 16802, USA}

\author{Ciar\'{a}n Hickey}
\affiliation{School of Physics, University College Dublin, Belfield, Dublin 4, Ireland}
\affiliation{Centre for Quantum Engineering, Science, and Technology, University College Dublin, Dublin 4, Ireland}

\author{Romain Vasseur}
\affiliation{Department of Theoretical Physics, University of Geneva, 1211 Geneva, Switzerland}
\affiliation{Geneva Quantum Center, University of Geneva, Geneva, Switzerland}

\author{Justin H.~Wilson}
\email{jhwilson@lsu.edu}
\affiliation{Department of Physics and Astronomy, Louisiana State University, Baton Rouge, Louisiana 70803, USA}
\affiliation{Center for Computation and Technology, Louisiana State University, Baton Rouge, Louisiana 70803, USA}

\begin{abstract}
Information-theoretic phase transitions, such as the measurement-induced phase transition (MIPT), characterize the robustness of quantum dynamics to local monitoring and are naturally formulated in terms of trajectories conditioned on typical measurement outcomes, which are naively accessible only through postselection.
Here, we implement forced measurements to investigate how explicit postselection alters the nature of the transition.
We find that postselection fundamentally alters the universality class by reweighting trajectories that are otherwise rare.
In particular, we obtain a correlation-length exponent $\nu\approx 2.1$ larger than that of the standard MIPT and a negative effective central charge $c_\mathrm{eff}\approx -0.35$.
We also compare the postselected MIPT to the entanglement transition of random tensor networks, and demonstrate that their universality class is the same.
This setup further allows time-periodic, translationally invariant circuits with postselected weak measurements.
In both models, we find that an on-site dimension of at least 3 (qutrits but not qubits) is necessary to induce a transition.
\end{abstract}

\maketitle

\emph{Introduction}---Many-body quantum systems under unitary time evolution can be characterized by a growth of entanglement and scrambling of local information into nonlocal degrees of freedom when the dynamics is sufficiently ergodic.
However, this growth can be slowed down and halted when local degrees of freedom are \emph{monitored},
leading to a measurement-induced phase transition (MIPT) from an \emph{entangling phase} to a \emph{disentangling phase}~\cite{skinnerMIPT19,Li2018,Chan2019,Choi2020,Gullans2020,Bao_stat_mech_2020,Jian_stat_mech_2020,Li2021,zabalo2022,Huse2020,Tang2020,Turkeshi2020}.
These information-theoretic phase transitions can be considered from various perspectives~\cite{Gullans2020,Choi2020,PhysRevLett.129.200602,PRXQuantum.5.020304}.
A particularly insightful perspective is provided by starting from mixed states (rather than pure states) in the monitored circuit: in the area law phase, frequent measurements purify the state at a system-size independent rate, whereas in the volume law phase, the purification time diverges exponentially in the system size~\cite{Gullans2020}. 

A similar entanglement transition was identified prior to the study of measurement-induced phase transitions in random tensor networks (RTNs), where the tensors are sampled from a distribution \cite{PhysRevB.100.134203, levy2021entanglemententropytransitionsrandom, hayden_holographic_2016, PhysRevB.105.104306}. 
In these networks, the boundary state undergoes an analogous area-to-volume law transition as one tunes the bulk \textit{bond dimension}.
While superficially unrelated, one can understand both the MIPT and the RTN transition through a mapping to a replica stat-mech model, where the transition manifests as an ordinary symmetry breaking transition in the effective ``spin'' degrees of freedom \cite{PhysRevB.100.134203}.
The crucial difference in their stat-mech description stems solely from the difference in sampling procedure: in MIPTs, measurement outcomes are sampled according to the Born distribution, whereas in RTNs the tensors are independent identically distributed.
This stat-mech framework then leads to a rather nontrivial prediction: if one were to get rid of standard measurements that are governed by the Born rule and instead postselect to a fixed measurement outcome, the two transitions collapse to the same universality class.

\begin{figure*}[t]
    \centering    \includegraphics{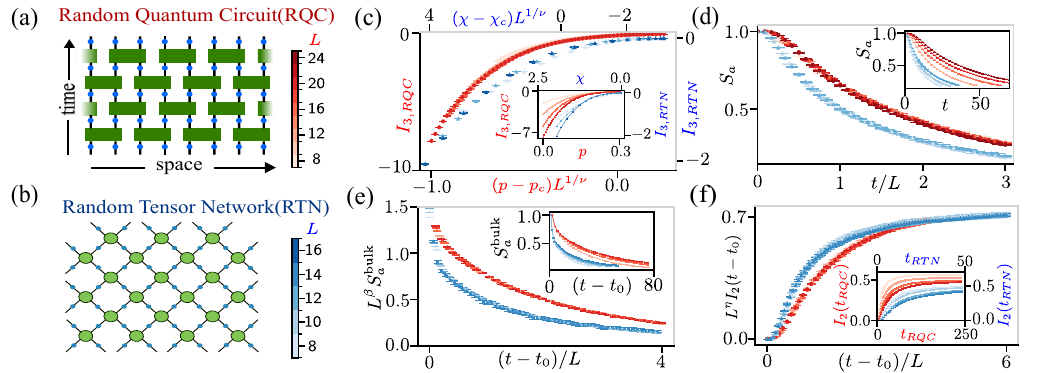}
    \caption{(a),(b) Schematic diagrams of both models: (a) circuit diagram of the random quantum circuit (RQC) and (b) circuit diagram of the random tensor network (RTN).
    (c)--(f) Scaling collapse of various entanglement measures.
    Different system sizes are shown in shades of blue (RTN) and red (RQC).
    (c) Tripartite mutual information, $I_3$, near the critical point $p_c(\chi_c)$ (\( L_\mathrm{RQC} = 8,12,16,20\) and \( L_\mathrm{RTN} = 8,12,16\)).
    (d) Time evolution of the von Neumann entropy of an ancilla maximally entangled with $L$ qubits (\( L_\mathrm{RQC} = 8,12,16,20,24\) and \( L_\mathrm{RTN} = 8,10,12\)).
    (e) von Neumann entropy of an ancilla entangled with a single qubit at $t > 4L$ (\( L_\mathrm{RQC} = 12,16,18,20\) and \( L_\mathrm{RTN} = 8,10,12,14\)).
    (f) Mutual information between ancillae entangled at sites $r$ and $r'$ such that $|r - r'| = L/2$ (\( L_\mathrm{RQC} = 8,12,16,20\) and \( L_\mathrm{RTN} = 8,10,12,14\)). For details on number of realizations, see End Matter.}
    \label{fig:schematics_and_results_app}
\end{figure*}

Postselection is already a heavily discussed topic in the literature on standard Born-rule MIPTs \cite{annurev:/content/journals/10.1146/annurev-conmatphys-031720-030658, Potter_2022, vasseur2026leshoucheslecturesrandom}.
The purported ``problem'' stems from the fact that MIPTs are transitions of quantum trajectories and therefore can only be detected using quantities that are nonlinear in the density matrix, in contrast to linear observables that can be characterized through the measurement-averaged density matrix.
As a result, one must repeatedly sample the same quantum trajectory to observe the transition, an event that is exponentially rare in the extensive number of measurements, making experimental detection rather challenging.
This issue has been partially mitigated for smaller system sizes using hybrid quantum circuits with cross-entropy benchmarking~\cite{noel_measurement-induced_2022,ExpMIPT25,google_measurement-induced_2023} and tomography techniques~\cite{koh_measurement-induced_2023}.
However, scaling to larger systems remains prohibitive due to the exponential growth of the Hilbert space, which necessitates either exponential computational resources for classical decoding or an exponential number of realizations of a quantum trajectory for reliable quantum state tomography.

Despite these difficulties, we now understand that MIPTs belong to a broader class of information-theoretic transitions, in which the two phases differ in how information spreads and whether it can be recovered from the system \cite{Choi2020,Gullans2020,PhysRevLett.129.200602,agrawal2023observingquantummeasurementcollapse,dehghani_neural-network_2023,PRXQuantum.5.020304}.
From this perspective, a postselected MIPT---where the model is defined \emph{ab initio} by conditioning measurements on a fixed outcome---should no longer be viewed as pathological, but instead as a natural realization of a distinct universality class of such information-theoretic transitions, one that has received comparatively little attention in the literature (exceptions include monitored fermions, see Refs.~\cite{PhysRevB.106.134206, jian2023measurementinducedentanglementtransitionsquantum, PhysRevX.13.041045, PhysRevX.13.041046, PhysRevResearch.6.043246}).
The connection with RTNs makes the postselected universality especially compelling since in RTNs the information-theoretic interpretation of the entanglement transition is natural: in the area-law phase, contracting the network is efficient, whereas in the volume-law phase it becomes exponentially hard in the system size.

In this Letter, we numerically study both postselected monitored quantum circuits and RTNs and characterize their critical properties.
In particular, we report several critical exponents and confirm that the two models fall into the same ``postselected'' universality class. 
We emphasize that here ``postselected'' refers to explicit postselection in the circuit dynamics, in which we fix the measurement outcomes of every circuit realization. 
This still suffers from the exponential overhead described previously for Born-rule MIPTs. To further probe the role of spatiotemporal randomness in these transitions, we reinstate spatial and time translational symmetries in both models in ways described in the main text.
The evolution is then effectively implemented by a single transfer matrix.
Interestingly, we observe a transition in the resulting RTN with three-state legs and in qutrit circuits, but not for qubits---raising the question of whether randomness within individual realizations is essential for the finite-$d$ monitored criticality.

\emph{Models}---The postselected random quantum circuit (RQC) consists of two-qubit gates, \( U_{i,i+1} \), drawn from the Haar distribution and arranged in a one-dimensional brick-layer geometry~\cite{zabalo2020} with randomly interspersed forced measurements with probability $p$ that project the measured qubits onto \( \ket{0} \).
Figure~\ref{fig:schematics_and_results_app}(a) depicts the gates as green rectangles and the forced measurements as blue circles.
Numerically, we will consider systems of $L$ qubits with periodic boundary conditions and run the dynamics for a time $t \gtrsim 4L$ to reach a steady state unless otherwise specified.
The rate of postselected measurements $p$ is varied to probe the MIPT.
Formally, a state $\ket{\psi_t}$ at a discrete time step $t$ is evolved via
\[
 \ket{\psi_{t+1}} = \prod_{i} M_i \prod_{i, \text{odd}} U_{i,i+1}\prod_{i} M_i \prod_{i, \text{even}} U_{i,i+1} \ket{\psi_{t}},
\]
where $\ket{\psi_{t+1}}$ is the (unnormalized) state at time $t+1$, $M_i = P_0^{(i)}$ with (\emph{classical}) probability $p$ and $M_i = I$ with probability $1-p$; $P_0^{(i)}$ is the projection onto the $\ket{0}$ state of qubit $i$, and $I$ is the identity operator.
Note that the randomness in the model is entirely due to the random sampling of $M_i$ and $U_{i,i+1}$---there is thus no notion of quantum trajectories weighted by Born probabilities of different measurement outcomes.

An RTN is a two-dimensional network of local tensors $T_{\mu,\nu,\kappa,\lambda}$ [see Fig.~\ref{fig:schematics_and_results_app}(b)].
The indices $\mu,\nu,\kappa,\lambda = 1,\dots,D$ label states in the local \emph{qudit} Hilbert space $\ket{1},\dots,\ket{D}$, where $D$ is the bond dimension of the network.
The tensor entries are drawn independently from a Gaussian distribution with zero mean and unit variance.
By adding a two-legged tensor (or equivalently, a quantum state) $\lambda_{\mu,\nu}$ on each edge $e$, we continuously tune the effective bond dimension of each leg parametrized by the second R\'enyi mutual information, $\chi_{n=2}$, of the two-legged state $\ket{\lambda} = \sum_{\nu,\mu=1}^D \lambda_{\mu,\nu}\ket{\mu}\ket{\nu}$ across $e$.
We will be interested in the state of the tensor network obtained at the boundary after the full contraction of the tensor network in the $L \times t$ lattice with periodic boundary conditions with the initial layer contracted at $t=1$ with a product initial state $\ket{0}^{\otimes L}$.
We perform the contraction in a circuitlike fashion, as shown in Fig.~\ref{fig:schematics_and_results_app}(b).
Although this is not strictly necessary, it enables us to adapt well-established protocols for extracting critical data in MIPTs to the analogous setting of RTNs.
Accordingly, throughout the rest of this Letter, we refer explicitly only to MIPT protocols, with the understanding that corresponding techniques apply analogously to RTNs.
Calling the resulting contracted tensor $\mathcal T_{\mu_1, \dots, \mu_N}$, the (unnormalized) wave function is $\ket{\psi} = \sum_{\mu_1, \dots, \mu_N} \mathcal T_{\mu_1, \dots, \mu_N} \ket{\mu_1, \dots, \mu_N}$.
Reference~\cite{PhysRevB.100.134203} demonstrated that tuning the effective bond dimension $D_{\mathrm{eff}}$
drives an entanglement transition in the disorder averaged R\'enyi entanglement entropy of the \textit{normalized} $\ket{\psi}$---from an area-law phase at small $D_{\mathrm{eff}}$ to a volume-law phase at large $D_{\mathrm{eff}}$. 
For a more detailed description, see End Matter. 

\emph{Results}---At the entanglement transition, the entanglement behaves critically in the steady state of these models. 
To diagnose this, we first define the $n$th R\'enyi entropy $S_n(A) = [1/(1-n)] \ln \tr \rho_A^n$, where $\rho_A = \tr_{\bar{A}} \ket{\psi}\bra{\psi}$, and compute the \emph{tripartite mutual information} (\(I_{3,n}\)) on four contiguous regions $A$, $B$, $C$, and $D$ of size $L/4$ \cite{zabalo2020}:
For the von Neumann case, we write \(I_3 \equiv I_{3,1}\),
\begin{multline}
    I_{3,n}(A, B, C) \equiv S_n(A) + S_n(B) + S_n(C) - S_n(A \cup B) \\
                     \quad - S_n(A \cup C) - S_n(B \cup C) + S_n(A \cup B \cup C).
    \label{eq:I_3}
\end{multline}

We locate the critical point \(p_c (\chi_c)\) by identifying the measurement rate (bond dimension) where the behavior of \(I_{3,n}\) changes.
In the area-law phase, \( I_{3,n} \) approaches zero for large \( L \) due to the cancellation of boundary contributions.
In contrast, in the volume-law phase, \( I_{3,n} \) exhibits a system-size dependence due to a nonzero bulk contribution~\cite{zabalo2020}, and at the critical point \( I_{3,n} \) is scale invariant.
From the exact numerical simulations of the RQC and the RTN, we use the scaling relation \( I_{3,1} \sim f[L^{1/\nu}(p-p_c)]\) to extract the critical points to be \( p_{c,\mathrm{RQC}} = 0.24(1) \) and \( \chi_{c,\mathrm{RTN}} = 0.90(9) \) and the correlation-length exponents to be \(\nu_{\mathrm{RQC}} = 2.1(4)\) and \(\nu_{\mathrm{RTN}} = 2.2(5)\) for the two models.

Coupling an ancilla locally with a system qubit by initializing them in a Bell state provides a local order parameter for purification \cite{Huse2020, Gullans2020}.
We obtain the density matrix of the ancilla $\rho_a$ by tracing out the system and we can compute $S_a = - \tr \rho_a\log \rho_{a}$ to diagnose purification.
We consider this probe in three contexts.
(1) We maximally entangle the ancilla with all qubits [by initializing the system in the state, \(\ket{\psi} = (1/\sqrt{2})(\ket{A\uparrow} + \ket{B\downarrow})\), where \(\ket{A}\) and \(\ket{B}\) are orthogonal states of the system drawn from the Haar distribution], then run the dynamics, tracking $S_a$ to diagnose purification.
Using \(S_a\), scaling yields \(p_c = 0.22(2)\) and \(\chi_c = 0.84(2)\), which agrees with the critical point obtained from \(I_3\) (see End Matter for details).
Applying the scaling form \(S_a \sim g(t / L^z)\) to the time evolution of \(S_a\) at the critical point, we obtain the dynamical exponents \(z_{\mathrm{RQC}} = 0.96(5)\) and \(z_{\mathrm{RTN}} = 1.05(5)\) as shown in \cref{fig:schematics_and_results_app}(d).
(2) We run the full dynamics to a time $t_0>4L$ (steady state) and couple in the ancilla locally in a Bell state to measure the local order parameter, \(S_a^\mathrm{bulk}\).
Using \(S_a^\mathrm{bulk}\) and the scaling relation $S_a^\mathrm{bulk} \sim L^{-\beta} h\!\left((t - t_0)/L^z\right)$, we extract the critical exponents $\beta_{\mathrm{RQC}} = 0.16(5)$ and $\beta_{\mathrm{RTN}} \approx 0.16(2)$ as shown in \cref{fig:schematics_and_results_app}(e).
We also extracted the critical exponent of the Born-rule MIPT to be \(\beta_\mathrm{B\text{-}RQC} = 0.10(4)\).
(3) We couple in two ancillae $a_1$ and $a_2$ into Bell states at either different locations or times and compute bipartite mutual information $I_{2}=S_{a_1}+S_{a_2}-S_{a_1a_2}$ to diagnose correlation functions.
In this scenario, the ancillae are coupled in at a time $t_0>4L$ in the steady state (bulk) and then evolved in time. 
Using \(I_{2}\), we extract the spatial correlation exponent, finding \( \eta_{\mathrm{RQC}} = 0.17(2) \) and \( \eta_{\mathrm{RTN}} = 0.16(2) \), as shown in \cref{fig:schematics_and_results_app}(f). 

\begin{small}
\begin{table}[t]
    \centering
    \begin{tabular}{c @{\hspace{0.2cm}} c @{\hspace{0.2cm}} c @{\hspace{0.2cm}} c @{\hspace{0.2cm}} c @{\hspace{0.2cm}} c@{\hspace{0.2cm}} c}
        \hline
        \hline
        Model & $p_c$ & $\nu$ & $\eta$ & $z$ & $\beta$ & $c_{\mathrm{eff}}$ \\
        \hline
        P-RQC & $0.24(1)$ & $2.1(4)$ & $0.17(2)$ & $0.96(5)$ & $0.16(5)$ & $-0.35(7)$ \\
        RTN & $0.90(9)$ & $2.2(5)$ & $0.16(2)$ & $1.05(5)$ & $0.16(2)$ & $-0.35(2)$\\
        B-RQC & $0.168(5)$ & $1.2(2)$ & $0.19(1)$ & $1.06(4)$ & $0.10(4)$& $0.25(3)$\\
        P & $0.5$ & $1.33$ & $0.21$ & $1$ & $0.139$ & $0.2914$\\
        \hline
        \hline
    \end{tabular}
    \caption{Critical exponents for the phase transition in the postselected random quantum circuit (P-RQC), the random tensor network (RTN), the Born-weighted random quantum circuit (B-RQC) \cite{zabalo2020, zabalo2022}, and the percolation transition (P) \cite{zabalo2022}.}
    \label{tab:critical_exponents}
\end{table}
\end{small}

Next, for both the postselected MIPT and the RTN, the mapping to a statistical-mechanical model implicitly defines a partition function $Z_X=\tr[\rho_X]$, where $\rho_X=\ket{\psi_X}\bra{\psi_X}$ is the unnormalized pure-state density matrix obtained by evolving an initial state under a disorder realization $X$ to a final state $\ket{\psi_X}$ \cite{PhysRevLett.99.120601}.
(Note that we use the term ``disorder'' here to denote random realizations of the circuits or tensors in the respective models, rather than quenched spatial randomness.)
At sufficiently long times, any dependence on the initial state can be neglected.
In the postselected MIPT, $X$ specifies the measurement locations and gates, while in the RTN it specifies the realization of the random tensors.
Equivalently, $Z_X$ is just the squared norm of the boundary state obtained after evolution with disorder realization $X$.
The squared norm plays a crucial role here: each disorder realization carries its own normalization. To average over these weights, one introduces the replica partition function $Z_k=\overline{[Z_X]^k}$ and the associated \textit{annealed} free energy $F_k=-\log Z_k$, where the overbar denotes the disorder average over $X$.
The \textit{quenched} average free energy is then $F=\lim_{k\to 0} dF_k/dk = -\overline{\log[Z_X]}$, which is the quantity of interest for the postselected MIPT and RTN.

For comparison, in Born-rule MIPTs one must also average over measurement trajectories $\mathbf{m}$ for a fixed $X$.
The replica partition function then becomes $\tilde{Z}_N=\sum_\mathbf{m} p_\mathbf{m}Z_\mathbf{m}^k=\sum_\mathbf{m} Z_\mathbf{m}^N$, where $N=k+1$, $p_\mathbf{m}=Z_\mathbf{m}$, and where the average over disorder $X$ is now implicit.
This extra replica changes the replica limit: the analogous free energy is $\tilde{F}=\lim_{N\to 1} d\tilde{F}_N/dN = -\sum_\mathbf{m} p_\mathbf{m}\log p_\mathbf{m}$, with $\tilde{F}_N=-\log \tilde{Z}_N$, so the limit is now $N\to 1$ rather than $0$.
For fixed gates and measurement locations, this reduces to the Shannon entropy of the measurement record. Additionally, note that the role of normalization in Born-rule MIPTs is different from the RTNs and postselected MIPTs considered in this Letter. In the former, for each disorder realization $X$ of gates and measurement locations, the norm $Z_{\mathbf m}[X]$ is the Born probability $p_\mathbf{m}$ of the measurement record $\mathbf m$, and therefore obeys $\sum_{\mathbf m}Z_{\mathbf m}[X]=1$ for each $X$. In contrast, the norm $Z[X]$ for the systems considered here plays no role in the sampling of the tensors or circuits.

Now consider a system of size $L$ with periodic boundary conditions, evolved for $t\gg L$.
The associated statistical-mechanical model lives on a cylinder of circumference $L$ and height $t$.
Conformal invariance then fixes the finite-size scaling of the quenched free energy, more precisely its density, to take the form \cite{LUDWIG1987687, LYKKEJACOBSEN1998701}
\begin{figure}[t]
    \centering
    \includegraphics[width=\columnwidth]{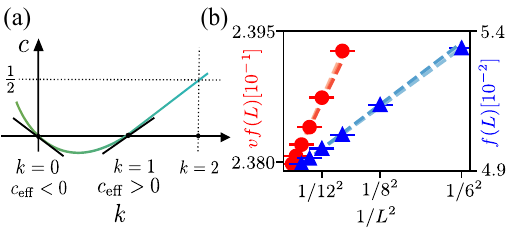}
    \caption{(a) Schematic of the central charge $c(k)$ of the replica partition function $Z_k$. $k=0,1$ correspond to the postselected and Born-rule MIPTs, respectively, and $k=2$ is the Ising model, which has $c=1/2$. (b) Free-energy density $f(L)$ [see Eq.~\eqref{eq:casimir}] vs $1/L^2$ (red circles, RQC; blue triangles, RTN).} 
    \label{cefffig}
\end{figure}
\begin{equation}
   \frac{F(L,t)}{A} = f(L) = f(L= \infty) - \frac{\pi c_{\mathrm{eff}}}{6L^2},
    \label{eq:casimir}
\end{equation}
with the space-time area \(A \equiv vLt\), where \(v\) is the anisotropy factor calculated by comparing the temporal and spatial two-point correlation functions at the critical point \cite{zabalo2022} (see End Matter for details).
Here, \(c_{\mathrm{eff}}\) is a universal number called the effective central charge, which is given by \(c_{\mathrm{eff}} = \lim_{k\rightarrow0}\left(dc(k)/dk\right)\), with $c(k)$ the central charge of the conformal field theory describing the transition in the statistical model at a finite number of replicas with free energy $F_k = -\ln Z_k$.
The anisotropy factor for the RQC is computed to be \(v_{\mathrm{RQC}} = 0.8(1)\).
As the RTN is symmetric under the exchange of space and time by construction, its anisotropy factor is \(v_{\mathrm{RTN}} = 1\).
Fitting the free energy density, $F(L,t)/A = f(L)$, to the Casimir form in \cref{eq:casimir} yields the \(c_{\mathrm{eff}}\) for the RQC and RTN models (see End Matter for details).
We numerically estimate the effective central charges for the models to be \(c_{\mathrm{eff,RQC}} = -0.35(7)\) and \(c_{\mathrm{eff,RTN}} = -0.35(2)\). 
The negativity of $c_{\rm eff}$, although atypical, follows from three facts~\cite{patil2025c}, namely, (i) \( c(k) = 0 \) at both \( k = 0 \) and \( k = 1 \) due to trivial partition functions in these limits; (ii) for the Born-rule MIPT, the effective central charge at \( k = 1 \), defined as \(c_{\mathrm{eff}}(k=1) = \lim_{k \to 1} [d c(k)/d k]\), is positive~\cite{zabalo2022}; and assuming (iii) \( c(k) \) is a continuous function of \( k \) without additional zeroes \footnote{The assumption of no zeros for $0<k<1$ is rather natural (although not rigorous) since $c=0$ theories with trivial partition functions usually arise due to some special structure in the problem and are not accidental. Besides the RTN and Born-rule MIPTs, physical examples of such $c=0$ theories include percolation ($q\rightarrow 1$ limit of the $q$-state Potts model) and self-avoiding walks [$n\rightarrow 0$ limit of the $O(n)$ model], where again, the emergence of $c=0$ is tied to the specific structure of the problem.}. 
Together, (i)--(iii) imply that the slope of \( c(k) \) at \( k = 0 \), i.e., \(c_{\mathrm{eff}}(k=0) = \lim_{k \to 0} [d c(k)/d k]\), must be negative as depicted in \cref{cefffig}.

\begin{figure}[t]
    \centering
    \includegraphics{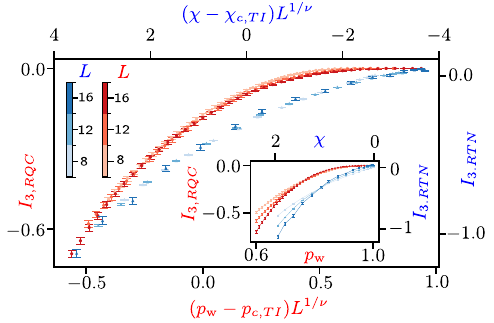}
    \caption{Tripartite mutual information $I_3$ for translationally invariant postselected weak-measurement variants (\( L_\mathrm{RQC} = 8,12,16\) and \( L_\mathrm{RTN} = 8,12,16\)).}
    \label{fig:ti_i3}
\end{figure}

\emph{Translationally invariant postselected weak measurements}---Postselection, or equivalently removing randomness from the measurement trajectories, naturally prompts the question of what happens when all remaining randomness is eliminated, including that in the gates and measurement locations. 
The resulting evolution is governed by a single transfer matrix built from local nonunitary gates.
One might naively expect this nonunitary evolution to mimic imaginary-time evolution toward the ground state of a local Hamiltonian with low entanglement, but it is unclear whether this intuition holds generically (see Refs.~\cite{wang2025stabilityquantumchaosweak, PhysRevLett.126.170503, PhysRevB.109.235102} for related works).
To investigate this, we now turn to time-periodic, translationally invariant systems where a single two-site unitary (or local tensor) is drawn once and repeated throughout the space-time lattice.
Interestingly, for the translationally invariant RTN with three-state physical legs, we find an entanglement transition as we tune the effective bond dimension ($\chi$), captured by $I_3$ as shown in \cref{fig:ti_i3}. 
The analogous fixed-unitary qubit circuit with postselected weak measurements, however, shows no transition in $I_{3,1}$ (see End Matter for details)---but moving to qutrits restores the transition as the weak-measurement strength $p_{\mathrm{w}}$ is varied. 
This contrast suggests a strong dependence on the on-site Hilbert-space dimension; in particular, the qutrit transition indicates that randomness within individual realizations is not strictly necessary for MIPT criticality in these postselected weak-measurement settings.

\emph{Conclusion}---We have characterized the universality of the postselected MIPT and demonstrated its relation to the RTN model.
The major differences in critical properties as compared to the standard MIPT (see Table~\ref{tab:critical_exponents}) originate from a reweighting of trajectories (equivalently, the replica limit $k= 0$). Closely related forced-measurement replica limits have been studied in monitored free-fermion settings, where Born-rule and forced ensembles are described by distinct replica and sigma-model theories \cite{PhysRevB.106.134206, jian2023measurementinducedentanglementtransitionsquantum, PhysRevX.13.041045, PhysRevX.13.041046, PhysRevResearch.6.043246}; however, the phase transitions studied here belong to a postselected Haar universality class as opposed to the Anderson type criticality exhibited by free-fermion models \cite{PhysRevB.106.134206, jian2023measurementinducedentanglementtransitionsquantum,PhysRevX.13.041045, PhysRevX.13.041046,PhysRevResearch.6.043246}.
Furthermore, we find that the effective central charge is negative for this universality class, which is rather unusual for disordered systems. We leave the exploration of the interpolation in \cref{cefffig} as a framework for understanding partial postselection and hybrid Born-forced protocols to future work~\cite{PRXQuantum.5.020304, agrawal2023observingquantummeasurementcollapse, wang2025stabilityquantumchaosweak, PhysRevLett.126.170503}.
Finally, we have also studied time-periodic, translationally invariant variants with postselected weak measurements, finding a transition in the translationally invariant RTN with three-state legs and in qutrit circuits (see \cref{fig:ti_i3}).

While the relationship is striking, there are still a few puzzles.
For instance, in Fig.~\ref{fig:schematics_and_results_app}, the critical exponents are seen to match, but the scaling functions appear to differ.
We attribute this mismatch to nonuniversal corrections and/or finite-size effects that are not well captured at these system sizes. In any case, the agreement of the extracted critical exponents supports the identification of the two transitions with the same postselected universality class. 
Furthermore, our estimated $\nu\sim2$ is close to the Harris bound~\cite{Harris74,Chayes86}, but with an uncertainty that makes it unclear on which side of the bound it lies. This proximity however suggests an interesting direction of future study, especially when comparing with the Born-rule MIPT, which flows to infinite randomness criticality in the presence of quenched disorder \cite{infinite-randomness}.
Intriguingly, $p_c$ is larger after postselection than it was in the Born-rule MIPT, suggesting that even naive selection of trajectories could enhance the volume-law phase's ability to retain quantum information.
We leave it to future work to explore whether this can be generalized or is a special feature of this model.

\begin{acknowledgments}
We thank S. Gopalakrishnan and A. Ludwig for helpful discussions.  
J.H.W. acknowledges support from the National Science Foundation (NSF) under Grant No.~DMR-2238895.
T.I. acknowledges support from NSF Grant No.~2611305. R.V. and K.K. are supported by the Swiss National Science Foundation (Grant No.~10008234) and by the Foundation for the University of Geneva.  
Portions of this research were conducted with high performance computational resources provided by Louisiana State University (\href{http://www.hpc.lsu.edu}{http://www.hpc.lsu.edu}).
This research was done using services provided by the OSG Consortium, which is supported by the National Science Foundation Awards No.~2030508 and No.~2323298.
\end{acknowledgments}

\emph{Data availability}---The data that support the findings of this article are openly available~\cite{nambi_2026_code,nambi_2026_postselected}.

\newpage

\appendix

\emph{Random tensor network}---To set notation, we briefly introduce the random tensor network (RTN) model \cite{PhysRevB.100.134203} shown in \cref{fig:schematics_and_results_app}(b).
It is defined on a square lattice of size $L \times t$ with periodic boundary conditions, where each vertex-edge pair $(v,e)$ hosts a Hilbert space $\mathcal{H}_{ve}$ with basis states $\ket{\mu_{ve}} = \ket{1}, \dots, \ket{D}$.
Accordingly, one can associate to each bulk vertex $v$ the state  
\begin{equation}
        \ket{T_v} = \sum_{\mu_{ve_1}, \dots, \mu_{ve_z} = 1}^D 
        T_{\mu_{ve_1}, \dots, \mu_{ve_z}} 
        \ket{\mu_{ve_1}, \dots, \mu_{ve_z}},
\end{equation}
where $z=4$ is the coordination number [see Fig.~\ref{fig:schematics_and_results_app}(b)], and $T_{\mu_{ve_1}, \dots, \mu_{ve_z}}$ are the entries of the tensor at vertex $v$.
These entries are sampled from a Gaussian distribution with zero mean, $\bar{T}_{\mu_{ve_1}, \dots, \mu_{ve_z}} = 0$, and unit variance, such that \(
\overline{T^*_{\mu_{ve_1}, \dots, \mu_{ve_z}} T_{\nu_{ve_1}, \dots, \nu_{ve_z}}} 
= \delta_{\mu_{ve_1}, \nu_{ve_1}} \dots \delta_{\mu_{ve_z}, \nu_{ve_z}}\).
At each edge, we define an entangled pair state,
\begin{equation}
    \ket{I_e} = \sum_{\mu_{ve}, \mu_{v'e}=1}^D \lambda_{\mu_{ve}, \mu_{v'e}} \ket{\mu_{ve}, \mu_{v'e}},
\end{equation}
where $(v, v')$ are neighboring vertices connected by edge $e$. It is this state that is used to contract neighboring tensors. In other words, it implements the contraction $\sum_{\mu_{ve},\mu_{v'e}} T_{\mu_{ve},\dots}\lambda_{\mu_{ve},\mu_{v'e}}T_{\mu_{v'e},\dots}$ at each edge $e$.
The matrix $\lambda$ controls the entanglement in $\ket{I_e}$, and accordingly the effective bond dimension along edge $e$ that we parametrize via the $n$th R\'enyi mutual information,
\begin{equation}
\label{chineqn}
    \chi^{(n)}_{e} = \frac{2}{1-n} \log \tr (\lambda \lambda^{\dag})^n,
\end{equation}
which ranges from $0$ for a product state to $2 \log D$ for a maximally entangled $D$-dimensional Bell state with $\lambda_{ve,v'e} = D^{-1/2} \delta_{ve,v'e}$.
Thus, tuning $\lambda$ continuously from a product state to a Bell-state interpolates the effective bond dimension between $0$ and $D$. While the specific choice of this interpolation is not unique, the only requirement is to have a monotonically increasing $\chi_e^{(n)}$ as one tunes $\lambda$ from a product to a Bell state. Then, for a given $\lambda$, we are interested in the properties of the state $\ket{\psi}$ obtained by contracting the tensor network,
\begin{equation}
    \ket{\psi} = \bigotimes_{v} \bigotimes_{e} \bra{T_v} {I_e} \rangle,
\end{equation}
where $\ket{\psi}$ resides on the physical legs at the top boundary of the network.

For numerical results presented in this Letter, we always fix the bare Hilbert-space dimension on every tensor leg to be $D = 3$ (qutrits). For each disorder realization, the entries of $\ket{T_v}$ are independently drawn from the Gaussian ensemble described above. Thus, in the RTN calculations presented here, the only randomness comes from entries of $\ket{T_v}$, while $\lambda$ is fixed uniformly throughout the network in each realization. The physical tuning parameter used in the main text and figures is the second R\'enyi mutual information $\chi_e^{(2)}$, which depends on $\lambda$ according to Eq.~\eqref{chineqn} and increases monotonically from $0$ to $2\log 3$ as we interpolate $\lambda$. The RTN transition is therefore labeled by a critical value $\chi_c$, in direct analogy with the critical measurement rate $p_c$ in the postselected quantum circuit. However, $\chi$ is not to be viewed as a measurement probability; it quantifies the entanglement at the bonds of the tensor network and defines an effective bond dimension that we tune to access the RTN transition. 

\emph{Half-cut entropy}---To probe the entanglement structure of the system, we divide it into two equal partitions and compute the R\'enyi entropy of the subsystem.
In the volume-law phase, \(S_{n,1/2} \propto L\) (volume of the subsystem); in the area-law phase, \(S_{n,1/2} \propto O(1)\) (boundary of the subsystems); and at the critical point, \(S_{n,1/2} \propto \alpha(n) \log L\).

\begin{figure}[htp!]
    \centering    
    \includegraphics[width=0.6\linewidth]{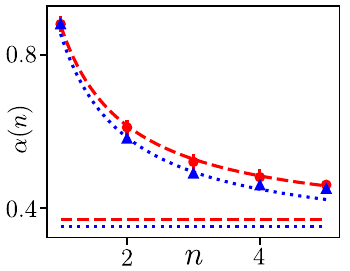}
    \caption{Coefficients \(\alpha(n)\) of the R\'enyi entropies (blue triangles, RTN; red circles, RQC).
    Blue dotted and red dashed lines show the \(n\) dependence, and the horizontal lines denote \(\alpha(\infty)\).}
    \label{fig:renyi}
\end{figure}

\begin{small}
\begin{table}[!ht]
    \centering
    \begin{tabular}{c @{\hspace{0.15cm}} c @{\hspace{0.15cm}} c @{\hspace{0.15cm}} c @{\hspace{0.15cm}} c @{\hspace{0.15cm}} c@{\hspace{0.15cm}} c}
        \hline
        \hline
        Model & $\alpha(1)$ & $\alpha(2)$ & $\alpha(3)$ & $\alpha(4)$ & $\alpha(5)$ & $\alpha(\infty)$ \\
        \hline
        P-RQC & $0.88(2)$ & $0.61(2)$ & $0.52(2)$ & $0.48(2)$ & $0.46(1)$ & $0.37(2)$ \\
        RTN & $0.88(1)$ & $0.581(2)$ & $0.490(8)$ & $0.459(7)$ & $0.45(1)$ & $0.35(1) $\\
        B-RQC & $1.7(2)$ & $1.2(2)$ & \ldots  & \ldots & $0.9(1)$ & $0.7(1)$\\
        \hline
        \hline
    \end{tabular}
    \caption{Coefficients $\alpha(n)$ of the half-cut R\'enyi entropies at criticality for the P-RQC, the RTN, and the B-RQC~\cite{zabalo2020, zabalo2022}.}
    \label{tab:halfcut_critical_exponents}
\end{table}
\end{small}

The critical exponents \(\alpha(n)\) for the random quantum circuit and random tensor network are consistent with each other and differ from the Born-rule MIPT, as shown in \cref{fig:renyi} and \cref{tab:halfcut_critical_exponents}.
For B-RQC, the ellipses indicate values not explicitly tabulated in Refs.~\cite{zabalo2020, zabalo2022}.
Their \(n\) dependence is \(\alpha_\mathrm{P\text{-}RQC}(n) = 0.53(3)/n  + 0.35(7)\) and \(\alpha_\mathrm{RTN}(n) = 0.55(2)/n + 0.32(4)\), in contrast to \(\alpha_\mathrm{B\text{-}RQC}(n) = 1.0(1)/n + 0.7(1)\).

\begin{figure}[htp!]
    \centering
    \includegraphics[width = 0.9\linewidth]{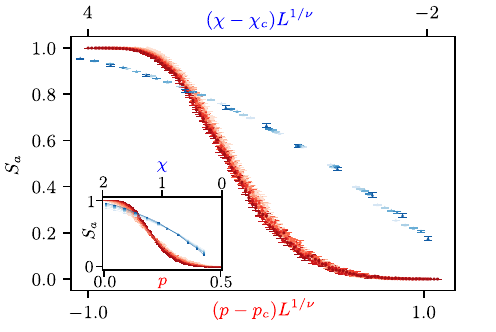}
    \caption{Scaling collapse of the ancilla entropy \(S_a\) near the critical point [\(p_c\) for RQC (red circles), \(\chi_c\) for RTN (blue triangles), \( L_\mathrm{RQC} = 8,10,12,14,16,18,20\) and \( L_\mathrm{RTN} = 8, 10,12,14, 16\)].}
    \label{fig:ancilla_p}
\end{figure}

\emph{Purification transition}---The measurement-induced phase transition can also be viewed as a purification transition.
To probe this, we maximally entangle an ancilla with the system in a Bell state and compute its steady-state von Neumann entropy \(S_a\) for different measurement rates, extracting \(p_c\) and \(\nu\) from the scaling collapse in \cref{fig:ancilla_p}.

\emph{Anisotropy factor computation}---In order to extract the finite-size free-energy density of the stat-mech models, one must account for the possible anisotropy between space and time. This anisotropy is captured through the anisotropy factor and is evaluated by comparing correlation functions along spatial and temporal directions (see Ref.~\cite{zabalo2022} for details of the procedure). In generic random unitary circuits, this anisotropy is already present microscopically: the unitary gates \textit{a priori} define a temporal direction and are not, in general, unitary under a space-time exchange of indices. By contrast dual unitary circuits are indeed unitary under such an exchange of indices and have an anisotropy factor of 1. Likewise, the RTNs considered in this work also have the space-time symmetry by construction and have an anisotropy factor of unity.

\begin{figure}[htp!]
    \centering    \includegraphics[width=\linewidth]{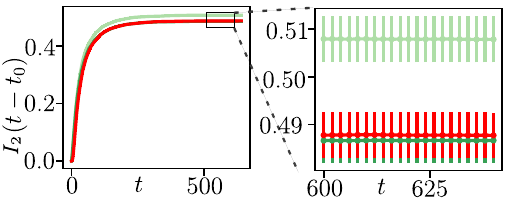}
    \caption{Green curves show temporal correlations for qubits separated by \(\Delta t = 7L/20\) and \(6L/20\) (dark to light); the red curve shows the spatial correlation for \(\Delta r = L/2\).}
    \label{fig:placeholder}
\end{figure}

Following the procedure used for the spatial correlation in the main text, the temporal correlation is obtained by entangling two ancilla qubits with the same qubit at times $(r,t)$ and $(r,t')$.
For $L=20$, the spatial correlation curve lies between the temporal curves for $t-t' = 6L/20$ and $7L/20$ as shown in \cref{fig:placeholder}, giving \(t^* \approx 6.9L/20\) by linear interpolation and the anisotropy factor \(v = \ln(1+\sqrt{2})(L/\pi t^*) = 0.8(1)\). The same analysis for \(L=16\) gives us \(v = 0.70(6)\) and \(c_\mathrm{eff} = -0.40(6)\) [which is within error bars of \(c_\mathrm{eff}\) computed using \(v = 0.8(1)\)].

\begin{figure}[ht]
    \centering
    \includegraphics[width=0.7\linewidth]{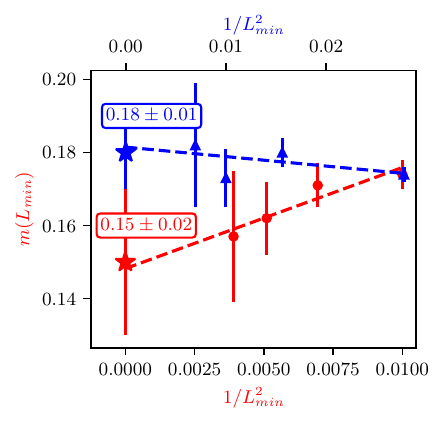}
\caption{Slope $m(L_{\mathrm{min}})$ as a function of $1/L_{\mathrm{min}}^2$ (blue triangles, RTN; red circles, RQC). The extrapolated values $m(\infty)$ obtained from linear fits are indicated with stars at $1/L_{\mathrm{min}}^2=0$, with the corresponding numerical values shown in the labeled boxes.}
    \label{fig:mfit}
\end{figure}

\emph{Effective central charge}---The effective central charge of both the random quantum circuit and the random tensor network is obtained by fitting the finite-size scaling of the free energy \(F\) (see main text) to the Casimir form in \cref{eq:casimir}.
\cref{cefffig} shows the free energy as a function of \(1/L^2\). To reduce finite-size effects on the computation of \(c_\mathrm{eff}\), we successively remove smaller systems to compute the slope \(m(L_{\mathrm{min}})\) (\(-\pi vc_\mathrm{eff}/6\)) by fitting into the functional form in \cref{eq:casimir}.
We then extrapolate \(m(\infty)\) from \(m(L_{\mathrm{min}}) = m(\infty) + b/L_{\mathrm{min}}^2\), to obtain the \(c_\mathrm{eff}\) in the thermodynamic limit using the expression \(c_{\mathrm{eff}} = -6m(\infty)/\pi v\) as shown in \cref{fig:mfit}.
The error bars for the effective central charge \(\Delta c_\mathrm{eff} = |c_\mathrm{eff}|\sqrt{[\Delta m(\infty)/m(\infty)]^2 + (\Delta v/v)^2}\) is obtained by propagating the errors in \cref{eq:casimir}, where \(v\) is the anisotropy factor.

\begin{figure}[htp!]
    \centering \includegraphics{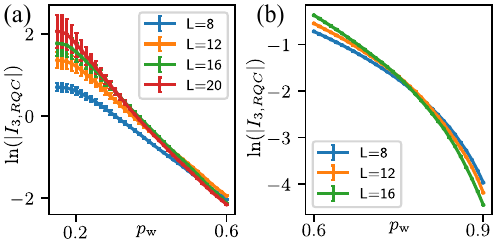}
    \caption{\(\ln |I_3|\) for the translationally invariant RQC: (a) qubits, (b) qutrits.}
    \label{fig:log_i_3}
\end{figure}

\begin{figure}[htp!]
    \centering
    \includegraphics[width=0.7\linewidth]{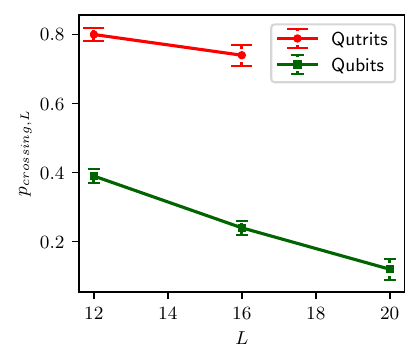}
    \caption{Qubits: Crossings are \(0.39(2)\) for \(L = 8,12\), \(0.24(2)\) for \(L = 12,16\) and \(0.12(2)\) for \(L = 16,20\). Qutrits: Crossings are \(0.80(2)\) for \(L = 8,12\) and \(0.74(3)\) for \(L = 12,16\).}
    \label{fig:pcdrift}
\end{figure}

\emph{Translationally invariant circuit}---As explained in the main text, we do not observe a MIPT for the translationally invariant RQC with qubits, but the transition reappears for qutrits.
For qubits, the weak-measurement crossover \(p_\mathrm{w}\) drifts to 0 with increasing system size; for qutrits it remains relatively stable, consistent with an MIPT in the thermodynamic limit, as shown in \cref{fig:log_i_3}.
The crossing of the curves corresponding to system sizes \(L-4\) and \(L\), \(p_{\mathrm{crossing},L}\), is plotted as a function of \(L\) in \cref{fig:pcdrift}. The qutrit crossing shifts by only \(7.5\%\) from \((8,12)\) to \((12,16)\), compared with \(38\%\) for the qubit crossings.

\emph{Number of realizations}---The number of independent realizations used to compute the entanglement measures for the RQC is \(4800\) for \(I_3\), \(S_a\), and \(S_a^{\mathrm{bulk}}\), \(9600\) for \(I_2(t-t_0)\), \(160\,000\) for $f(L)$ for all system sizes shown in \cref{fig:schematics_and_results_app,cefffig} and \(400\) for \(I_3\) in \cref{fig:ti_i3}. For the RTN, the number of realizations depends on both the system size $L$ and a little on the bond dimension \(\chi\). For \(I_3\) in \cref{fig:schematics_and_results_app}, we use \(15\,000\) realizations for (\(L=8\) and \(12\)), and \(1542\)--\(2048\) realizations for (\(L=16\)). For \(S_a\), we use \(1.5\times10^4\)--\(3.0\times10^4\) realizations for (\(L=8\), \(10\), and \(12\)), \(4.9\times10^3\)--\(1.8\times10^4\) realizations for (\(L=14\)), and \(5.1\times10^2\)--\(1.3\times10^3\) realizations for (\(L=16\)). For \(S_a^{\mathrm{bulk}}\), we use \(6000\) realizations for ($L=8$ and $10$), \(5068\) realizations for (\(L=12\)), and \(339\) realizations for (\(L=14\)). For \(I_2(t-t_0)\), we use \(10\,000\), \(12\,030\), and \(3707\) realizations for (\(L=8\), \(12\), and \(14\)), respectively. For the free-energy calculations \(f(L)\), we use \(50\,000\) realizations for (\(L=6\), \(8\), and \(10\)), \(150\,000\) for (\(L=12\)), \(26\,521\) for (\(L=14\)), and \(2517\) for (\(L=16\)). For \(I_3\) in \cref{fig:ti_i3}, we use \(40\,000\) realizations for (\(L=8\) and \(12\)), and \(1967\)--\(2636\) realizations for (\(L=16\)).

\end{document}